\documentclass[review]{elsarticle}

\usepackage{amssymb}
\usepackage{graphicx}
\usepackage{amsmath}
\usepackage{algorithm}
\usepackage{algpseudocode}
\usepackage{bm}
\usepackage{url}
\usepackage{amsfonts}
\usepackage{indentfirst}
\usepackage{booktabs}
\usepackage{diagbox}
\usepackage{graphicx}
\usepackage{color}
\usepackage{tabularx}
\usepackage{multirow}
\usepackage{booktabs}
\usepackage{float}
\usepackage{marvosym}
\usepackage{enumerate}
\usepackage{enumitem}
\usepackage{graphicx}
\usepackage{subfloat}
\usepackage{subfigure}

\newtheorem{theorem}{Theorem}

\newtheorem{remark}{Remark}

\algrenewcommand\algorithmicrequire{\textbf{Input:}}
\algrenewcommand\algorithmicensure{\textbf{Output:}}

\usepackage{lineno,hyperref}
\modulolinenumbers[5]

\begin{document}

\begin{frontmatter}

\title{On Rangasamy's outsourcing algorithm for solving
	quadratic congruence equations\tnoteref{mytitlenote}}
\tnotetext[mytitlenote]{This research is supported by National Key Research and Development
	Program of China (No. 2020YFA0712300), National
	Natural Science Foundation of China (No. 62032009).}

\author[mymainaddress,mysecondaryaddress]{Xiulan Li}
\author[nankai]{Yansong Feng}
\author[mymainaddress,mysecondaryaddress]{Yanbin Pan \corref{mycorrespondingauthor}}
\cortext[mycorrespondingauthor]{Corresponding author}
\ead{panyanbin@amss.ac.cn}

\address[mymainaddress]{Key Laboratory of Mathematics
	Mechanization, Academy of Mathematics and Systems Science, Chinese
	Academy of Sciences, Beijing 100190, China}
\address[mysecondaryaddress]{School of Mathematical
Sciences, University of Chinese Academy of Sciences, Beijing 100049, China}
\address[nankai]{School of Mathematical Sciences, Nankai University, Tianjin, 300071, China}

\begin{abstract}
Outsourcing computation is a desired approach for IoT (Internet of Things) devices to transfer their burdens of heavy computations to those nearby, resource-abundant cloud servers. 
Recently, Rangasamy presented a passive attack against two outsourcing algorithms proposed by Zhang et al. for  solving quadratic congruence equations, which is widely used in IoT applications. Furthermore, he also proposed a modified algorithm to fix these schemes and claimed that his algorithm was correct and enabled secure and
verifiable delegation of solving quadratic congruence equations in IoTs. However, we show that Rangasamy's modified  algorithm has a flaw which makes it incorrect and also propose some further attacks to break the security claim, even when 
 the flaw has been corrected. 
\end{abstract}


\end{frontmatter}


\section{Introduction}\label{sec:intro} 
The internet of things (IoT), which is one of emerging technologies in the Fourth Industrial Revolution, is likely to play a dominant role in what emerges post-pandemic \cite{GTGR}. Information is collected, stored and shared across the internet by networked `smart' physical objects, such as RFID tags, sensors, mobile phones etc. By 2025, it is predicted that 41.6 billion devices will be capturing data on how we live, work, move through our cities, operate and maintain the machines on which we depend according to the World Economic Forum's State of the Connected World report \cite{SoCW}. However, almost all of these deployed devices have limited computing and storage capacity. As a result, outsourcing computation is a desired approach for these devices to transfer their burdens of heavy computations to those nearby, resource-abundant cloud servers by pay-as-you-go model.

With the rapid development of cloud computing, a large number of cloud servers offering computing and storage make it possible to securely outsource computation tasks. However, for the sake of business interests, equipment breakdown and etc., there are security risks in outsourcing computation. So it must meet three requirements, which are high efficiency, input/output privacy and verifiability \cite{DBLP:conf/infocom/HuAACSL17}.
 
Solving quadratic congruence equations, which is to find root of $x^2\equiv a\bmod n$ given a quadratic residue $a$ modulo $n$, is  widely used in cryptographic constructions, such as Rabin Cryptosystem \cite{DBLP:books/cu/Galbraith2012}. Meanwhile, this public-key encryption scheme is suitable in IoT applications. However, it needs time $O(\log^3p)$ for the quadratic congruence problem with prime modulus $p$, which makes the computation overloaded for those IoT devices. 

Therefore, many scholars have been studying how to securely outsource the problem of solving quadratic congruence equations to cloud servers. Recently, Zhang et al. \cite{ZYT2020} proposed two outsourcing algorithms, SoSQC1 and SoSQC2, for this problem, and claimed that all the original inputs and output cannot be exposed by the cloud servers in their algorithms. However, Li et al. \cite{FixedSoSQC}  presented some attacks to show that the  SoSQC1 and SoSQC2 schemes are insecure since all the inputs and outputs can be efficiently recovered by  just a curious server. They also presented another algorithm to fix these two algorithms.  In another independent work,   Rangasamy \cite{Rangasamy21} also  questioned the security of  SoSQC1 and SoSQC2 schemes and presented  a passive attack to recover the inputs and output when the protocol is executed more than once. Furthermore,  Rangasamy also proposed a modified algorithm and claimed that his new algorithm was correct and enabled secure and
verifiable delegation of solving quadratic congruence equations in IoTs.

However, in this note, we show that the modified outsourcing algorithm proposed by Rangasamy is incorrect. The legal client usually cannot get the correct answer for his outsourcing task when completing the modified algorithm with an honest server. Furthermore, we show that Rangasamy's modified algorithm is insecure even when the flaw has been corrected. By proposing some attacks inspired by  \cite{FixedSoSQC}, we show that a curious server can successfully recover the original inputs and real output of the outsourcing task, which should be kept secret.

\section{Description of Rangasamy's modified algorithm}\label{sec:des}
In this section, we describe Rangasamy's modified algorithm \cite{Rangasamy21}, which fixes Zhang $et\;al.$'s algorithm SoSQC2 \cite{ZYT2020}. 
 Rangasamy's algorithm is also based on Cipolla's algorithm, which is usually used to find a solution of the quadratic congruence equation $x^2\equiv n \bmod p$, and  shown as Algorithm \ref{Alg:Copolla}.

\begin{algorithm}[H]
	\caption{Cipolla's Algorithm}
	\label{Alg:Copolla}
	\begin{algorithmic}[1]
		\Require  an odd prime $p$, and a quadratic residue $n\in \mathbb{F}_p$.
		\Ensure $x,$ such that $x^2\equiv n \bmod p$.
		\State  Find an $a\in \mathbb{F}_p$ until
		$
		\left(a^{2}-n\right)^{\frac{p-1}{2}} \equiv -1 \bmod p,
		$
		witch means that $a^2-n$ is a quadratic nonresidue modulo $p$.
		\State    Compute the root 
$
		x\equiv (a+\sqrt{w})^{\frac{p+1}{2}} \bmod p
$
		in the field $\mathbb{F}_{p^2}=\mathbb{F}_p(\sqrt{w})$, where $w=a^2-n$. 
		\State	\Return $x$.
	\end{algorithmic}
\end{algorithm}

Next we describe Rangasamy's algorithm as follows, in which the client outsources the task of solving quadratic congruence equation $x^2\equiv n \bmod p$ to the server. 

\begin{enumerate}
	\item The client randomly  picks a large prime $q$, whose bit-length is the same as that of $p$ and random integers $r_1, r_2, k$ ($k$ must  be small enough to ensure the efficiency) in $\mathbb{F}_p$, and then computes:
	\begin{eqnarray}
	&&n^{\prime}=n-r_1p, \label{nprime}\\
	&&d^{\prime}=(p-1)/2-k, \label{dprime}\\
	&&d_2^{\prime}=\frac{p+1}{2}+r_2(p-1), \label{d2prime} \\
	&&p^{\prime}=pq. \label{pprime}
	\end{eqnarray}	
	Then the client sends $(n^{\prime},d^{\prime},d_2^{\prime},p^{\prime})$ to the server.
	\item The server selects integer $a\in \mathbb{F}_{p^{\prime}}$ and calculates  $R_1^{\prime}\equiv (a^2-n^{\prime})^{d^{\prime}}\mod p^{\prime}$.  Then  $(a, R_1^{\prime})$ is sent back to the client.
	\item The client computes $R_1\equiv R_1^{\prime} \bmod p$ and checks whether or not $(a^2-n^{\prime})^k\cdot R_1\equiv -1 \bmod p$. If so, a message ``Y" is returned to the server; otherwise, ``N" is returned.
	\item Upon receiving ``N", the server repeats Step 2 by selecting another $a$ until receiving ``Y". Then the server calculates
	\begin{equation}\label{r2prime}
	R_2^{\prime}\equiv (a+\sqrt{a^2-n^{\prime}})^{d_2^{\prime}} \bmod p^{\prime},
	\end{equation}
	and sends $R_2^{\prime}$ to the client.
	\item The client computes 
	\begin{equation}
	\label{realx}
	x\equiv R_2^{\prime} \bmod p,
	\end{equation} 
	and checks whether or not $x^2 \equiv n \bmod p$.
\end{enumerate} 
  
\section{ Rangasamy's algorithm is incorrect}\label{sec:cor}

Unfortunately, we have to say that  Rangasamy's modified algorithm is incorrect, that is, even when the algorithm is executed honestly, the result $x$ computed by the client in Equation (\ref{realx}) cannot pass validation for correctness. 

Denote $w=a^2-n$. Note that
$ R_2^{\prime}\bmod p\equiv (a+\sqrt{w})^{\frac{p+1}{2}+r_2(p-1)} \bmod p$  computed in  Equation (\ref{realx}) may not be equal to $(a+\sqrt{w})^{\frac{p+1}{2}} \bmod p$ that is desired by Cippolla's algorithm (Step 2 in Algorithm 1), since  $R_2^{\prime}\bmod p \in \mathbb{F}_p(\sqrt{w})$ may not fall into $\mathbb{F}_p$ due to the fact that  the order of the multiplicative group in $\mathbb{F}_p(\sqrt{w})$ is $p^2-1$ instead of $p-1$. Hence the correctness of Rangasamy's algorithm does not hold. 

In the following, we take outsourcing $x^2\equiv 9 \bmod 83$ as a counterexample.  The client first randomly chooses  $q=97, r_1=21, r_2=73, k=13$,  computes
\begin{eqnarray*}
&&n^{\prime}=n-r_1p=-1734,\\
&&d^{\prime}=(p-1)/2-k=28,\\
&&d_2^{\prime}=\frac{p+1}{2}+r_2(p-1)=6028, \\
&&p^{\prime}=pq=8051,
\end{eqnarray*}
 and sends them to the cloud server. 

On receiving parameters $(n^{\prime},d^{\prime},d_2^{\prime},p^{\prime})$, the cloud server selects integer $a=3345$ and sends to the client the value $$R_1^{\prime}\equiv (a^2-n^{\prime})^{d^{\prime}}\mod p^{\prime}\equiv 11190759^{28}\bmod 8051\equiv 3927.$$

The client checks  $$(a^2-n^{\prime})^k\cdot R_1^{\prime}\equiv11190759^{13}\cdot3927\bmod 83 \equiv -1,$$ and return ``Y" to the server.

 Receiving ``Y", the cloud server computes $$R_2^{\prime}\equiv (a+\sqrt{a^2-n^{\prime}})^{d_2^{\prime}} \bmod p^{\prime}\equiv (3345+\sqrt{11190759})^{6028}\bmod 8051\equiv 3935\sqrt{7920}+5592$$ and sends $3935\sqrt{7920}+5592$ to the client. 

 Finally, the client computes $$x\equiv R_2^{\prime}\bmod p\equiv 34\sqrt{35}+31$$ and finds that $x^2\equiv n \bmod p$ doesn't hold since $x^2\equiv 33\sqrt{35} + 4$, which is not even an integer.

To ensure correctness, it seems one should at least set 
$$d_2^{\prime}=(p+1)/2+r_2(p^2-1)$$
by following Rangasamy's  idea. However, we next show that the algorithm is still insecure.

\section{Rangasamy's algorithm is insecure} \label{sec:anal}

The input/output privacy requires that the outsourcing algorithm should keep the original input $n$, $p$ and the correct output $x$ secret to anyone except the client. However, we next show that for Rangasamy's modified algorithm, these can be recovered efficiently by a curious server or an eavesdropper.  

Note that if we could recover $p$, then  we can easily recover $n\equiv n^{\prime} \mod p$ by Equation (\ref{nprime}) and recover $x$ by  solving the quadratic congruence equation. Therefore we just show how to recover $p$ in the following.

\subsection{Recovering $p$ from $d_2'$}
	By Equation (\ref{d2prime}), we  can get $2d_2^{\prime}-2=(p-1)(1+2r_2)$
which is exactly a multiple of $p-1$.  Moreover, this still holds even we set  $d_2^{\prime}=(p+1)/2+r_2(p^2-1)$, since now we have $2d_2^{\prime}-2=(p-1)(2r_2(p+1)+1).$

Note that the order of the cyclic multiplicative group $\mathbb{F}_p^{\ast}$ is $p-1$  since $p$ is a prime. Hence, for any $b\in \mathbb{F}_p^{\ast}$, we have $b^{2d_2^{\prime}-2}\equiv 1 \mod p$, from which we can get $p|(b^{2d_2^{\prime}-2}-1)$. With the fact that $p|p'$, we immediately  gets   
$$ p| g=\gcd((b^{2d_2^{\prime}-2}-1),p') = \gcd((b^{2d_2^{\prime}-2}-1)\bmod p',p').$$
If $b^{2d_2^{\prime}-2}-1\bmod p'\not\equiv 0$, then $p=g$ since the positive factors of $p'$ are in the set $\{1, p, q, p'\}$.	

From this observation above, the curious adversary can choose random integer $b$ until $b^{2d_2^{\prime}-2}-1 \mod p'\not\equiv 0$. Then he can immediately obtain the secret modulus $p$ by calculating $g$.	Due to the randomness of $r_2$ and $b$, the probability of $b^{2d_2^{\prime}-2}-1 \mod p'\not\equiv 0$ is very high.

We  randomly generated 100  instances on personal laptop to verify the effect of our attack, in which the bit lengths of randomly chosen $p$, $q$ are 512 bits and the bit length of $k$ is 80. In our experiments, we successfully recovered $p$ with all instances, that is, the success probability is 100\%.


\begin{remark}
	In fact, if the client executes the modified algorithm to solve $x_1^2\equiv n_1 \bmod p$ and $x_2^2\equiv n_2 \bmod p$ respectively  as assumed in \cite{Rangasamy21},  then the adversary knows queries $(n_1^{\prime},d^{\prime},d_2^{\prime},p^{\prime})$ and $(n_2^{\prime},\bar{d^{\prime}},\bar{d_2^{\prime}},{p^{\prime}})$, and
	 Rangasamy's idea \cite{Rangasamy21} can be directly employed to attack his modified algorithm, since  $2d_2^{\prime}-2$ and $2\bar{d_2^{\prime}}-2$ are the multiples of $p-1$ in the above two executions and  $\gcd(2d_2^{\prime}-2, 2\bar{d_2^{\prime}}-2)$ will leak $p-1$ with high probability. An analysis  similar to that in  \cite{Rangasamy21} shows that asymptotically the probability should be at least greater than 81.1\%, the probability  that two "random" odd integer are coprime.

	 We also generated 100 random instances when the client  executes the modified algorithm twice to verify the effect of this attack. The parameters are set as in our attack. Finally, we successfully recovered $p$ with probability 93\%.

\end{remark}

%
%

\subsection{A simple attempt to change $d_2'$ again}	
Based on the attacks above, we should force $2d_2'-2$ to be not a multiple of $p-1$. A simple idea to fix it is to substitute previous $d_2^{\prime}=(p+1)/2+r_2(p^2-1)$ with $d_2^{\prime}=(p+1)/2+r_2(p^2-1)-k_1$ with small $k_1$.

However, we have to point out that we still should be careful with the choice of $r_2$ in such a case. Again assume the adversary knows queries $(n_1^{\prime},d^{\prime},d_2^{\prime},p^{\prime})$ and $(n_2^{\prime},\bar{d^{\prime}},\bar{d_2^{\prime}},{p^{\prime}})$
 as in \cite{Rangasamy21}, where $d_2^{\prime}=(p+1)/2+r_2(p^2-1)-k_1$ and 
 $\bar{d_2^{\prime}}=(p+1)/2+\bar{r}_2(p^2-1)-\bar{k}_1$.
 If $r_2$ and $\bar{r}_2$ are small enough, saying less than $p$, then we have
 $$|\frac{d_2^{\prime}}{\bar{d_2^{\prime}}}-\frac{r_2}{\bar{r}_2}|< \frac{1}{O(p^2)},  $$
 and  continued fractions method \cite{HG01} may be an effective way to obtain $r_2$ and  $\bar{r}_2$ since $\frac{r_2}{\bar{r}_2}$ may be a best rational approximation of $\frac{d_2^{\prime}}{\bar{d_2^{\prime}}}$.

For example, suppose $p=691$ and we generate $d_2^{\prime}=325641678$ where  $r_2=682$, $k_1 = 28$ and $\bar{d_2^{\prime}}=313704683$ where $\bar{r}_2=657$,  $\bar{k}_1=23$ in the first and second executions, respectively. We computed the best rational approximations of $\frac{d_2^{\prime}}{d_2^{\prime\prime}}=\frac{325641678}{313704683}$ by the continued fractions method with  Sagemath \cite{sagemath} and get  the sequence $[1,
27/26,
82/79,
109/105,
191/184,
682/657,
28153/27121,
28835/27778,\\
114658/110455,
4271181/4114613,
34284106/33027359,
72839393/70169331,\\
325641678/313704683]
$ which contains the real $\frac{r_2}{\bar{r}_2}=\frac{682}{657}$.

After recovering $r_2$, we can recover $p$ from   $d_2^{\prime}=(p+1)/2+r_2(p^2-1)-k_1$ by the method similar to that in the following section when  $k_1$ is small enough.

\subsection{Recovering $p$ from $d'$ with small $k$}

Even $d_2'$ can be fixed in the modified algorithm, we have to show that it is still insecure with small $k$ in Equation \ref{dprime}.

Note that in Step 3 of  Rangasamy's algorithm, the 
client  must check whether  $(a^2-n^{\prime})^k\cdot R_1\equiv -1 \bmod p$ or not, which means that $k$ can not be too large since a large $k$ will cost the client too much resource. Usually the bit length of $k$ is set to be 80 to ensure 80-bit security as in \cite{ZYT2020}. However, small $k$ will lead some risk to leak $p$.

From Equation (\ref{dprime}), we can get $d^{\prime }=(p-1)/2-k$ for some small $k$ in $\mathbb{F}_p$. Thus, $k$ is the root of $g(x)\equiv 2x+2d^{\prime }+1-p^{\prime } \bmod p$ where $p^{\prime}=pq$. Since 2 is coprime to odd $p^{\prime}$, we can define another polynomial $f(x)=x+d^{\prime}+\frac{1-p^{\prime}}{2}$. It is apparent that  $k$ is also the root of $f(x) \bmod p$. Then we can recover $k$ by Coppersmith's algorithm in polynomial time if $k\leq\sqrt{p}$. More precisely, we have
\begin{theorem}[Coppersmith algorithm  \cite{isc10}] \label{th:Coppersmith}
	Let $f(x)$ be a univariate monic polynomial of degree $\delta$, and
	$N$ be an integer with unknown factorization. Assume that $N$ has a divisor $b\geq N^{\beta}$, where $0<\beta\leq 1$.  Then all solutions $x_0$ for the equation $f(x) \equiv 0 \mod b$ with $|x_0| \leq cN^{\frac{\beta^2}{\delta}}$ can be found
	in time $O(c\delta^5 \log^9 N)$.
\end{theorem}
Taking $p'$ as $N$, $p$ as $b$ in the theorem, we can get $\beta\approx \frac{1}{2}$ and $\delta= 1$ for $f(x)$ and then the bound $\sqrt{p}$ holds. 
Once $k$ is gotten, $p$ can be efficiently recovered from $d'$.

To validate the effectiveness of our attack,we randomly generated 100 instances, in which  $p$ and $q$ are  512 bits and $k$ is 80 bits. We succeeded in recovering $k$ for all the experiments. Moreover, we also tested for the case when  $p$, $q$ are 1024 bits and $k$ is 256 bits. 100 random instances were generated and we  succeeded in all the experiments again.


\section{Conclution} \label{sec:con}
In this note, we show that Rangasamy's modified outsourcing algorithm for  solving quadratic congruence equations has a flaw. Moreover, we present some attacks against it to show that all the inputs and output can be recovered efficiently, which breaks the security claim. We suggest the fixed algorithm in \cite{FixedSoSQC} as a candidate secure outsourcing algorithm for  solving quadratic congruence equations.

\bibliography{mybib}

\begin{thebibliography}{10}
\expandafter\ifx\csname url\endcsname\relax
  \def\url#1{\texttt{#1}}\fi
\expandafter\ifx\csname urlprefix\endcsname\relax\def\urlprefix{URL }\fi
\expandafter\ifx\csname href\endcsname\relax
  \def\href#1#2{#2} \def\path#1{#1}\fi

\bibitem{GTGR}
W.~E. Forum,
  \href{https://www3.weforum.org/docs/WEF_Global_Technology_Governance_2020.pdf}{Global
  technology governance report 2021: Harnessing fourth industrial revolution
  technologies in a covid-19 world}.
\newline\urlprefix\url{https://www3.weforum.org/docs/WEF_Global_Technology_Governance_2020.pdf}

\bibitem{SoCW}
W.~E. Forum,
  \href{https://www3.weforum.org/docs/WEF_The_State_of_the_Connected_World_2020.pdf}{State
  of the connected world}.
\newline\urlprefix\url{https://www3.weforum.org/docs/WEF_The_State_of_the_Connected_World_2020.pdf}

\bibitem{DBLP:conf/infocom/HuAACSL17}
C.~Hu, A.~Alhothaily, A.~Alrawais, X.~Cheng, C.~Sturtivant, H.~Liu,
  \href{https://doi.org/10.1109/INFOCOM.2017.8057199}{A secure and verifiable
  outsourcing scheme for matrix inverse computation}, in: 2017 {IEEE}
  Conference on Computer Communications, {INFOCOM} 2017, Atlanta, GA, USA, May
  1-4, 2017, {IEEE}, 2017, pp. 1--9.
\newblock \href {http://dx.doi.org/10.1109/INFOCOM.2017.8057199}
  {\path{doi:10.1109/INFOCOM.2017.8057199}}.
\newline\urlprefix\url{https://doi.org/10.1109/INFOCOM.2017.8057199}

\bibitem{DBLP:books/cu/Galbraith2012}
S.~D. Galbraith,
  \href{https://www.math.auckland.ac.nz/\%7Esgal018/crypto-book/crypto-book.html}{Mathematics
  of Public Key Cryptography}, Cambridge University Press, 2012.
\newline\urlprefix\url{https://www.math.auckland.ac.nz/\%7Esgal018/crypto-book/crypto-book.html}

\bibitem{ZYT2020}
H.~Zhang, J.~Yu, C.~Tian, G.~Xu, P.~Gao, J.~Lin,
  \href{https://doi.org/10.1109/JIOT.2020.2964015}{Practical and secure
  outsourcing algorithms for solving quadratic congruences in internet of
  things}, {IEEE} Internet Things J. 7~(4) (2020) 2968--2981.
\newblock \href {http://dx.doi.org/10.1109/JIOT.2020.2964015}
  {\path{doi:10.1109/JIOT.2020.2964015}}.
\newline\urlprefix\url{https://doi.org/10.1109/JIOT.2020.2964015}

\bibitem{FixedSoSQC}
X.~Li, J.~Bi, C.~Tian, H.~Zhang, J.~Yu, Y.~Pan, An improved outsourcing
  algorithm to solve quadratic congruence equations in internet of things,
  {IEEE} Internet Things J.\href {http://dx.doi.org/10.1109/JIOT.2021.3113013}
  {\path{doi:10.1109/JIOT.2021.3113013}}.

\bibitem{Rangasamy21}
J.~Rangasamy, \href{https://doi.org/10.1016/j.dam.2021.06.013}{On "practical
  and secure outsourcing algorithms for solving quadratic congruences in iots"
  from {IEEE} iot journal}, Discret. Appl. Math. 302 (2021) 139--146.
\newblock \href {http://dx.doi.org/10.1016/j.dam.2021.06.013}
  {\path{doi:10.1016/j.dam.2021.06.013}}.
\newline\urlprefix\url{https://doi.org/10.1016/j.dam.2021.06.013}

\bibitem{HG01}
N.~Howgrave{-}Graham,
  \href{https://doi.org/10.1007/3-540-44670-2\_6}{Approximate integer common
  divisors}, in: J.~H. Silverman (Ed.), Cryptography and Lattices,
  International Conference, CaLC 2001, Providence, RI, USA, March 29-30, 2001,
  Revised Papers, Vol. 2146 of Lecture Notes in Computer Science, Springer,
  2001, pp. 51--66.
\newblock \href {http://dx.doi.org/10.1007/3-540-44670-2\_6}
  {\path{doi:10.1007/3-540-44670-2\_6}}.
\newline\urlprefix\url{https://doi.org/10.1007/3-540-44670-2\_6}

\bibitem{sagemath}
{The Sage Developers}, {S}ageMath, the {S}age {M}athematics {S}oftware {S}ystem
  ({V}ersion 9.2), {\tt https://www.sagemath.org} (2020).

\bibitem{isc10}
A.~May, \href{https://doi.org/10.1007/978-3-642-02295-1\_10}{Using
  lll-reduction for solving {RSA} and factorization problems}, in: P.~Q.
  Nguyen, B.~Vall{\'{e}}e (Eds.), The {LLL} Algorithm - Survey and
  Applications, Information Security and Cryptography, Springer, 2010, pp.
  315--348.
\newblock \href {http://dx.doi.org/10.1007/978-3-642-02295-1\_10}
  {\path{doi:10.1007/978-3-642-02295-1\_10}}.
\newline\urlprefix\url{https://doi.org/10.1007/978-3-642-02295-1\_10}

\end{thebibliography}

\end{document}